# Video Coding for Machines:
# Partial transmission of SIFT features

Sławomir Maćkowiak, Marek Domanski, Sławomir Różek, Dominik Cywiński, Jakub Szkiełda

Poznań University of Technology, Poznań, Poland

*Abstract*— The paper deals with Video Coding for Machines that is a new paradigm in video coding related to consumption of decoded video by humans and machines. For such tasks, joint transmission of compressed video and features is considered. In this paper, we focus our considerations of features on SIFT keypoints. They can be extracted from the decoded video with losses in number of keypoints and their parameters as compared to the SIFT keypoints extracted from the original video. Such losses are studied for HEVC and VVC as functions of the quantization parameter and the bitrate. In the paper, we propose to transmit the residual feature data together with the compressed video. Therefore, even for strongly compressed video, the transmission of whole all SIFT keypoint information is avoided.

## I. Introduction

Traditional video coding algorithms, like AVC, HEVC, or VVC [16-19] were developed to provide the best subjective quality perceived by humans under certain bitrate conditions. However, in the last years, we observe rapid growth of machine vision applications. The proliferation of Internet of Things, autonomous vehicles, increasing employment of sensors, and the evolution of machine learning methods are only some of the factors that stimulate rash increase of amount of video data shared between computers, without direct human consumption. In response to the need of compressing these type of data, the MPEG sets up a Video Coding for Machines (VCM) Ad hoc Group, whose mission is to standardize a new codec optimized for consumption of decoded video by machines only or jointly by machines and humans [1-8]. For the prospective VCM codecs, MPEG VCM has specified preliminary requirements and use-cases related to Surveillance, Intelligent Transportation, Smart City, and Intelligent Industry [7].

As machines often operate on features extracted from images, the new video coding algorithm shall be capable of compressing both features and video. When performing machine tasks, the desired codec shall be more efficient than state-of-the-art techniques like VVC [18,19]. The bitstream shall be useful to perform various tasks on the decoder side, e.g. object detection, segmentation, or tracking (see Fig. 1). It shall be also suitable for devices with varying performance levels [7].

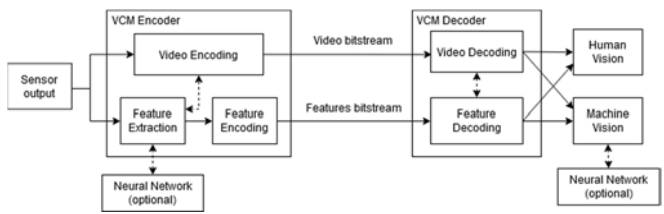

Fig. 1. Generic structure of encoders and decoders in VCM [1,7,8].

Significant efforts have been done for improving the efficiency of automatic analysis of decoded video, e.g. for object detection [3,4]. While VCM focuses on efficient compression of data for machine purposes, often it is also necessary to reproduce the image or video for both machine and human consumption. Hence, the VCM codec shall support data reconstruction for human inspection, which may require additional bitstream (Fig.1). Therefore, the joint compression of video and features is an interesting research topic [2,6]. Interesting results have been already obtained for joint coding of specific types of features like e.g. feature locations [12].

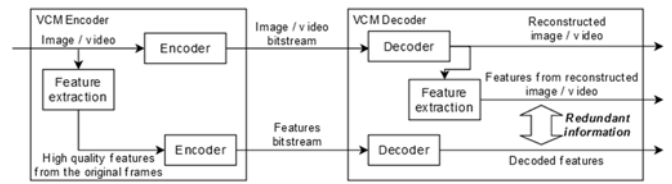

Fig. 2. The straightforward approach to coding of video and features.

The past research efforts were mostly focused on independent encoding of features or metadata. The developments include ISO MPEG-7 standard [10] on metadata and its compression. The newer parts of the MPEG-7 standard like Compact Descriptors for Visual Search (CDVS) [11] and Compact Descriptors for Video Analysis (CDVA) [13,14] have already gained some popularity and have proved their applicability. Unfortunately, these techniques have been designed for independent coding of metadata or features, but not for joint coding of video and features. Independent coding of video and features inevitably leads to certain redundancy, especially for higher bitrates (see Fig.2). In such an approach, features need to be completely transmitted in parallel to compressed video, despite of the fact that some subset of these features can be extracted from the decoded video. The goal of this paper is to study the aims to remove the abovementioned redundancy.

## II. The proposed approach to joint coding of video and features

We propose to extract features from both the original and the decoded video in the encoder. In that way only the difference between two sets of features needs to be transmitted (Fig.3). In the decoder, the full set of features is reconstructed as a sum of the sets of transmitted features and those extracted from the decoded video. Some features, e.g., keypoints may be lost due to video compression, and they should be send them as side information. Obviously, in many cases, the estimation of the difference or the sum of the feature is only a rough approximation of the operations needed. For some features, compression may change some of their parameters. Therefore, such parameters derived from the decoded video need to be replaced by the values transmitted as side information.

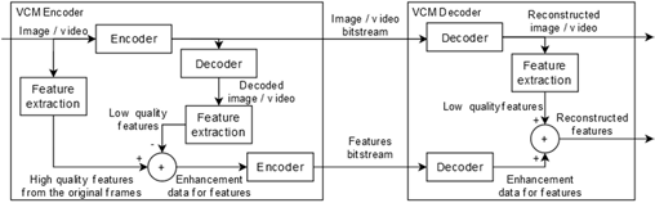

Fig. 3. The general structure of the proposed VCM encoder and decoder.

In this paper, we focus our considerations of video features on SIFT keypoints [9,22] as they are widely used (e.g. in the CDVS standard [11]). Similar problem has been already studied in [5] but using somewhat different approach.

## III. NUMBER OF SIFT KEYPOINTS EXTRACTED FROM VIDEO VERSUS VIDEO DEGRADATION DUE TO COMPRESSION

Firstly, we study the number of the SIFT keypoints that can extracted from video that was decoded with various values of quantization parameters (and bitrates of the compressed bitstream). Obviously, strong compression yield significant reduction of the number of SIFT keypoints that can be extracted from decoded video. Assuming, independent extraction from each video frame, we observe obvious fluctuations in the number keypoints obtainable from consecutive frames. The experimental results are obtained for two 1920×1080 MPEG surveillance test video sequences [15] (single views were extracted from these multiview clips). The plots represent the minimum, the average, and the maximum numbers of keypoints in a single frame from a 250-frame clip (Fig.5). We see that these variations are much smaller than the effects of compression with the quantization parameter QP ≥ 32 (for both HEVC and VVC compression). Also I-frames are slightly less prone to the keypoint losses, at least for higher bitrates (Fig. 5). The SIFT extractor from OpenCV (ver. 4.3.0) and the reference video codecs [20,21] were configured as in Common Test Conditions [24,25].

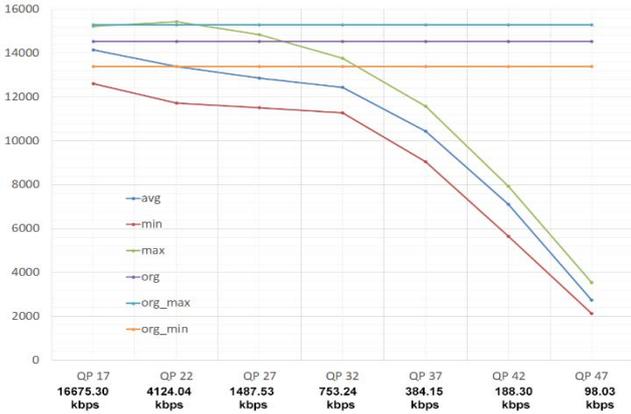

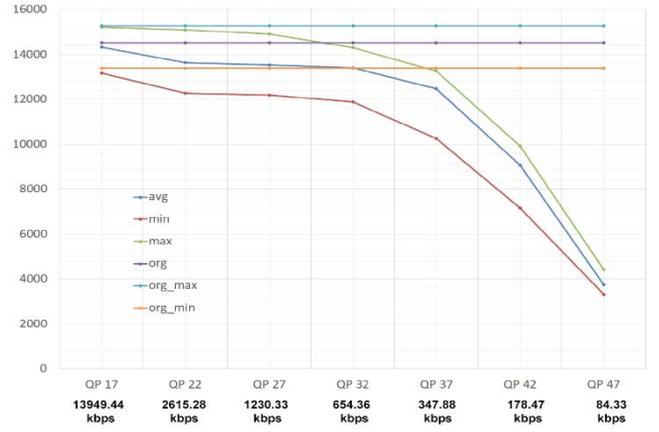

Fig. 4. Minimal, maximal and averaged (over 250 frames) number of keypoints per frame, Poznan CarPark sequence, HEVC compression (top), VVC compression (bottom). Horizontal lines indicate numbers of keypoints extracted from an uncompressed sequence.

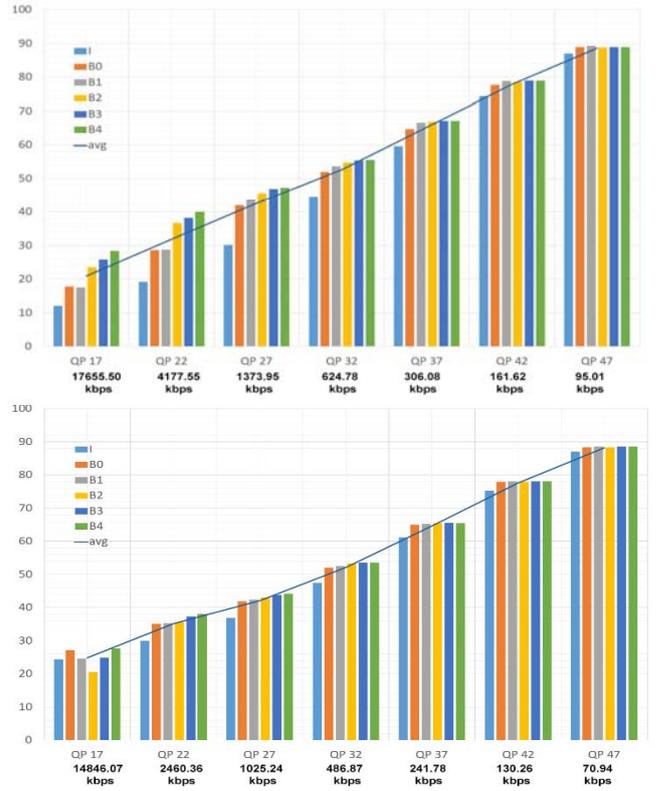

Fig. 5. The average count of SIFT keypoints that are not preserved in decoded video, i.e. that has vanished or are different in the sense of their locations by coding frame type and quantization parameter (HEVC encoding (top), VVC encoding (bottom), PoznańStreet sequence).

The total number of keypoints strongly depends on video content. Interestingly, the compression-implied relative loss of the keypoints (i.e. the percentage of all keypoints) seems to depend on the quantization parameters in a constant way independently from the video content and the codec type.

## IV. INFLUENCE OF COMPRESSION ONTO THE PARAMETERS OF THE SIFT KEYPOINTS EXTRACTED FROM DECODED VIDEO

Compression affects not only the number of keypoints but also the parameters of the SIFT keypoints extracted from decoded video. For each keypoint with position ($x,y$), the SIFT algorithm [9,22] provides the strength of the response to the presence of a corner and the dominant orientation based on two gradient parameters, and size parameter, which indicates the area coresponding to the keypoint. Therefore we consider 5 parameters of each keypoint: two coordinates of the location (*x*,*y*) and 3 other parameters: 'Response', 'Orientation', and 'Size'. In this section, we consider the influence of video compression onto those 5 parameters. For the sake of simplicity, we leave the effects related to SIFT descriptors beyond the scope of the considerations.

Let consider the influence of compression on keypoint locations (*x*,*y*). For the purpose of the study, the keypoint extracted from the original video are categorized into four groups with somewhat arbitrary definitions:
- *same* - keypoints of the original frame consistent in position with the keypoints extracted from the decoded frame. A point is considered the same if its position does not exceed one sampling point in one direction only.
- *moved* - keypoints shifted in the decoded image but within the boundary of the (7×7)-sample window around the location of the keypoints in the original frame,
- *missed* - keypoints not found in the decoded frame,
- *new* - keypoints present in the decoded frame that have no corresponding keypoint in the original frame.

The experimental results for selected test sequences are depicted in Fig. 6. For *QP* < 32, mostly the majority of keypoints belongs to the class 'same'. Moreover, the strength of influence is similar for different codecs for the same *QP* values.

In the next step, we consider the effects of compression for the remaining parameters. For the sake of brevity, we consider only the parameters of the keypoints from the class 'same', as for example. The considerations for the remaining classes would be similar with different numerical values.

In Fig. 7, the experimental results are depicted for two video test sequences and two video codecs, HEVC and VVC. The plots represent the numbers of keypoints with a given number of unchanged parameters, i.e. with their values within the limits of ± 5%. The curves are plotted as functions of the quantization parameter *QP*. For a given video test sequence, the curves look quite similarly, despite of the type of the codec used. The types of the curves with different value tolerances look similarly but they have to be omitted here.

V. PERFORMANCE OF THE SYSTEM

In the two previous sections, we have considered the numbers of keypoint parameters that need to be sent to the decoder because their values are significantly distorted by compression. We have estimated these numbers on the frame-by-frame as they would be estimated with no reference to previous frames. Obviously, such an approach is not efficient.

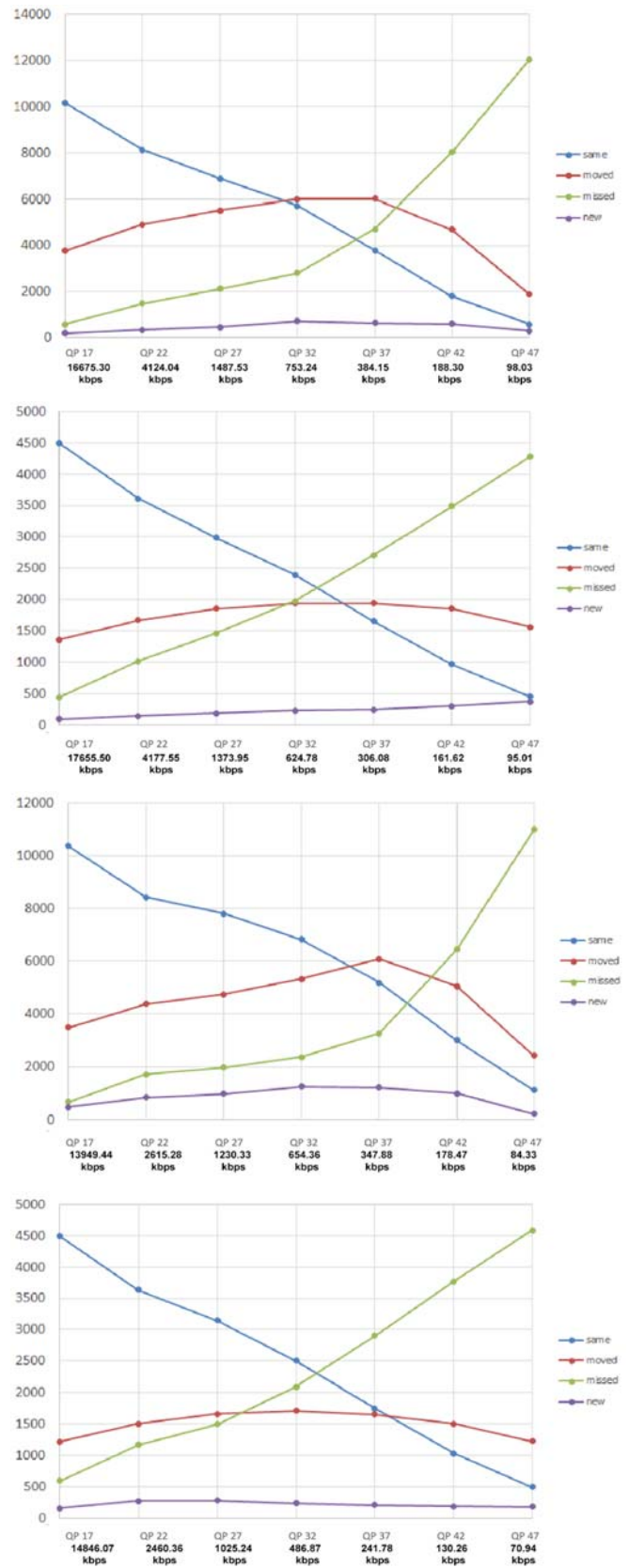

Fig.6. Examples of counts of keypoints in the categories, (from top to bottom respectively: HEVC encoding - Poznań CarPark, HEVC encoding - Poznań Street, VVC encoding - Poznań CarPark, VVC encoding - Poznań Street)

For efficient coding of the residual features, inter-frame prediction could be very efficient. Nevertheless, we have to leave such aspects for considerations elsewhere, e.g. in [5].

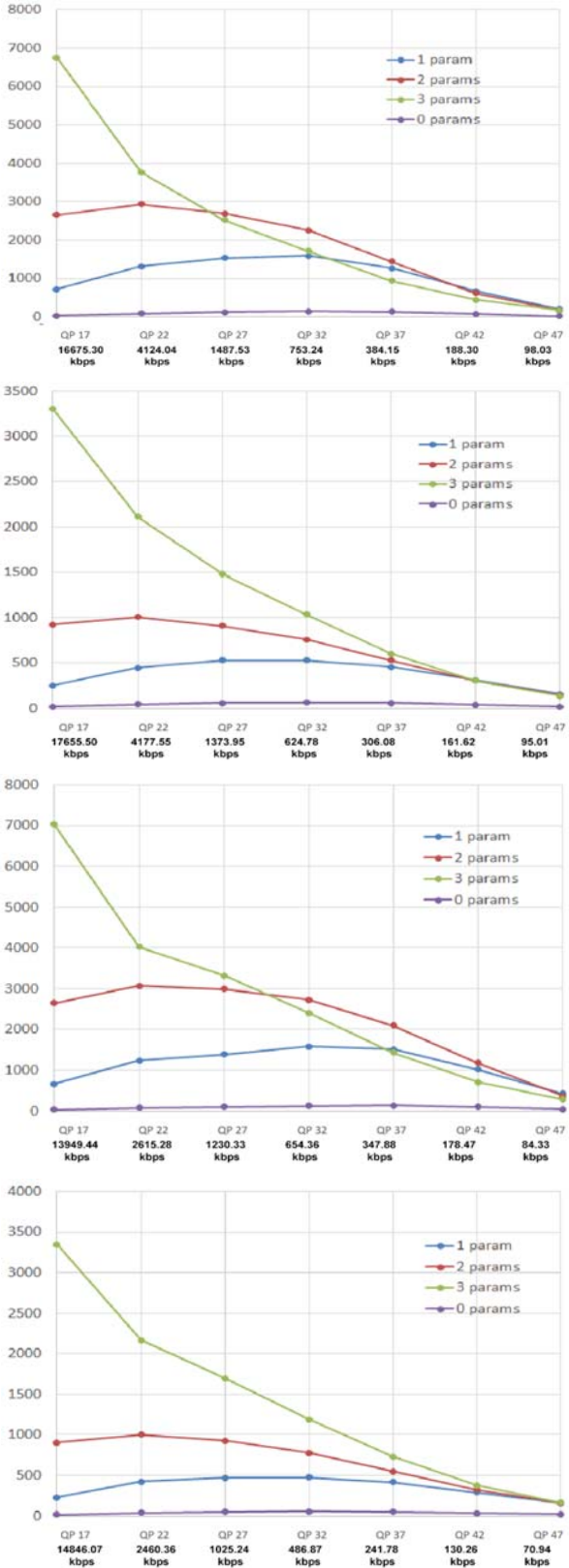

Fig.7. Number of keypoints with parameters within ±5 percent of the original keypoint parameter, (from top to bottom respectively: HEVC encoding - Poznań CarPark, HEVC encoding - Poznań Street, VVC encoding - Poznań CarPark, VVC encoding - Poznań Street).

Here, we use the results from the previous sections in order to estimate the amount of data transmitted as side information (feature bitstream) in the system shown in Fig.3. We also leave the important issue of compression (inter-frame prediction, entropy coding etc.) of this feature stream for further investigations.

Exploiting the results partially presented in the previous sections, we are able to plot an average number of the transmitted parameters per keypoint (Fig. 8). Again, the plot demonstrates that the keypoint data losses behave in the very similar way for both types of video encoders. Therefore, from the experimental data partially presented in the paper, we draw a rough formula

$$L\,[\%] \cong 24 + 1.4 \cdot QP\,. \qquad (1)$$

where $L$ is the percentage of the original parameters that have to be transmitted as side information in order to retrieve the original features in the decoder. The absolute errors of the formula are mostly in the range ±5%, at least for $QP$ in the range between 17 and 47.

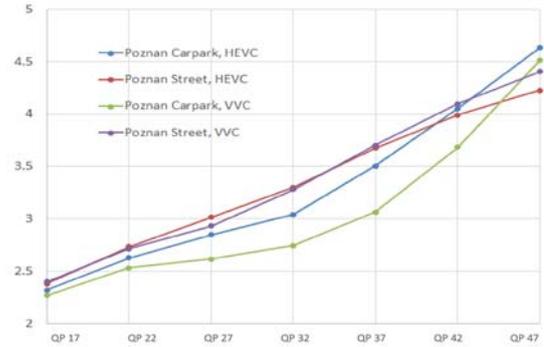

Fig. 8. Average number of keypoint parameters transmitted as side information assuming that the whole description is available in a decoder.

## VI. CONCLUSIONS

In the paper, we have proposed an original approach to Video Coding for Machines for joint coding of video together with SIFT keypoints and their parameters. The key issue is transmission of the residual keypoints that cannot be extracted from the decoded video after strong compression. For $QP \le 32$ (both HEVC and VVC), the number of such keypoints is moderate, whereas this number increases significantly for stronger compression. Moreover, as side information, transmitted are also some corrections to the keypoint parameters that are extractable from decoded video.

Astonishingly, the number of parameters that need to be transmitted as side information seems to be very similar for HEVC and VVC compression with the same value of the quantization parameter $QP$. The number of the keypoint parameters transmitted as side information is roughly proportional to the value of $QP$, according to Formula 1 derived in the paper. The formula is generic and seems to hold for both HEVC and VVC compression as well as for various video clips. Despite of the number of the experiments already done by the authors, the issue still needs further experimental validation using very different content.

Obviously the results of this paper remain valid also for still pictures.

ACKNOWLEDGMENT

The work was supported by Project xxx .

## REFERENCES

[1] L. Duan, J. Liu, W. Yang, T. Huang and W. Gao, "Video Coding for Machines: A Paradigm of Collaborative Compression and Intelligent Analytics," in IEEE Transactions on Image Processing, vol. 29, pp. 8680-8695, 2020.

[2] K. Fischer, F. Brand, C. Herglotz and A. Kaup, "Video Coding for Machines with Feature-Based Rate-Distortion Optimization," 2020 IEEE 22nd International Workshop on Multimedia Signal Processing (MMSP), 2020.

[3] L. Galteri, M. Bertini, L. Seidenari, and A. Del Bimbo, "Video compression for object detection algorithms," in Proc. International Conference on Pattern Recognition (ICPR), Aug. 2018, pp. 3007–3012.

[4] H. Choi and I. V. Bajic, "High efficiency compression for object detection," in Proc. IEEE International Conference on Acoustics, Speech and Signal Processing (ICASSP), Apr. 2018, pp. 1792–1796.

[5] J. Chao and E. Steinbach, "Keypoint encoding for improved feature extraction from compressed video at low bitrates," IEEE Trans. Multimedia, vol. 18, no. 1, pp. 25–39, Jan. 2016.

[6] S. Ma, X. Zhang, S. Wang, X. Zhang, C. Jia, and S. Wang, "Joint feature and texture coding: Toward smart video representation via frontend intelligence," IEEE Trans. Circuits Syst. Video Technol., vol. 29, no. 10, pp. 3095–3105, Oct. 2019.

[7] "Use cases and requirements for Video Coding for Machines," Doc. ISO/IEC JTC1/SC29/WG2 N18, October 2020.

[8] "Draft Evaluation Framework for Video Coding for Machines," Doc. ISO/IEC JTC1/SC29/WG2 N19, October 2020.

[9] Lowe D. G., Distinctive Image Features from Scale-Invariant Keypoints, International Journal of Computer Vision, 60(2), 2004, pp91-110.

[10] B.S. Manjunath, Ph. Salembier, Th. Sikora, "Introduction to MPEG-7, Multimedia content description interface," John Wiley & Sons, 2002.

[11] ISO/IEC 15938-13: Information Technology on Multimedia Content Description Interface, Part 13: Compact Descriptors for Visual Search, Sep. 2015.

[12] S. S. Tsai, D. Chen, G. Takacs, V. Chandrasekhar, M. Makar, R. Grzeszczuk, and B. Girod, "Improved coding for image feature location information," Proc. SPIE, vol. 8499, Oct. 2012, Art. no. 84991E.

[13] ISO/IEC 15938-15: Information Technology on Multimedia Content Description Interface, Part 15: Compact Descriptors for Video Analysis, Jul. 2019.

[14] L.-Y. Duan *et al.*, "Compact descriptors for video analysis: The emerging MPEG standard," IEEE Multimedia Mag., vol. 26, no. 2, pp. 44–54, Apr. 2019.

[15] M. Domański, T. Grajek, K. Klimaszewski, M. Kurc, O. Stankiewicz, J. Stankowski, K. Wegner, "Poznań multiview video test sequences and camera parameters", ISO/IEC JTC1/SC29/WG11 MPEG Doc. M17050, Xian, China, Oct. 2009.

[16] G. J. Sullivan, J. Ohm, W. J. Han, and T. Wiegand, "Overview of the High Efficiency Video Coding (HEVC) Standard", in IEEE Transactions on Circuits Systems for Video Technology, vol. 22, no. 12, pp. 1649-1668, Dec. 2012.

[17] ITU-T Rec. H.265 | ISO/IEC IS 23008-2, High efficiency coding and media delivery in heterogeneous environment – Part 2: High efficiency video coding.

[18] J. Chen, Y. Ye, S. Kim, "Algorithm description for Versatile Video Coding and Test Model 3 (VTM3)", Joint Video Experts Team (JVET) of ITU-T SG 16 WP 3 and ISO/IEC JTC 1/SC 29/WG 11, Doc. JVET-L1002, Macao, Oct 2018

[19] ISO/IEC DIS 23090-3 (2020) / ITU-T Recommendation H.266 (08/2020), "Versatile video coding".

[20] https://vcgit.hhi.fraunhofer.de/jct-vc/HM/-/tree/HM-16.20.

[21] https://vcgit.hhi.fraunhofer.de/jvet/VVCSoftware_VTM/-/tree/VTM-11.0.

[22] D. Lowe, "Object recognition from local scale-invariant features" Proceedings of the International Conference on Computer Vision. 2, 1999. pp. 1150–1157.

[23] D. Lowe, "Distinctive Image Features from Scale-Invariant Keypoints". International Journal of Computer Vision, vol. 60 (2), 2004, pp. 91–110.

[24] F. Bossen, "Common test conditions and software reference configurations," Joint Collaborative Team on Video Coding (JCT-VC) of ITUT SG16 WP3 and ISO. IEC JTC1/SC29/WG11, Doc. JCTVC-J1100, Stockholm, Sweden, Jul. 2012.

[25] F. Bossen, J. Boyce, W. Li, V. Seregin, K. Sühring, "JVET common test conditions and software reference configurations for SDR video," Joint Video Experts Team (JVET) of ITU-T SG 16 WP 3 and ISO/IEC JTC 1/SC 29/WG 11, Doc. JVET-N1010, Geneva, Switzerland, Mar. 2019.